\documentclass[twocolumn,secnumarabic,amssymb,preprintnumbers,showpacs,nobibnotes, aps, prd,floatfix]{revtex4-1}
\usepackage{subfigure}
\usepackage{amsmath}
\usepackage{amsfonts}
\usepackage{amssymb}
\usepackage[dvips]{graphicx}
\usepackage[usenames]{color}
\usepackage[applemac]{inputenc}
\usepackage[T1]{fontenc}

\usepackage{bbm}
\def\beq{\begin{equation}}
\def\eeq{\end{equation}}
\def\bea{\begin{eqnarray}}
\def\eea{\end{eqnarray}}
\def\ba{\begin{array}}
\def\ea{\end{array}}

\def\part{\partial}

\begin{document}

\preprint{UdeM-GPP-TH-10-190}
\preprint{arXiv:****.****[***]}

\title{Phase transitions in a gas of anyons}
\author{R. MacKenzie$^1$}
\author{F. Nebia-Rahal$^1$}
\author{M. B. Paranjape$^1$}
\author{J. Richer$^2$}

\affiliation{$^1$Groupe de physique des particules, D\'epartement de
physique, Universit\'e de Montr\'eal, C.P. 6128, Succ. Centre-ville,
Montr\'eal, Qu\'ebec, CANADA, H3C 3J7 }
\affiliation{$^2$ Réseau québecois de calcul de haute performance, DGTIC, Université de Montréal, C.P. 6128, Succ. Centre-ville, Montréal, Québec, Canada, H3C 3J7}
\begin{abstract}
We continue our numerical Monte Carlo simulation of a gas of closed loops on a 3 dimensional lattice,  however now in the presence of a topological term added to the action corresponding to the total linking number between the loops.  We compute the linking number using certain notions from knot theory.  Adding the topological term converts the particles into anyons.   Using the correspondence that the model is an effective theory that describes the  2+1-dimensional Abelian Higgs model in the asymptotic strong coupling regime, the topological linking number  simply corresponds to the addition to the action of the Chern-Simons term.   We find the following new results. The system continues to exhibit a phase transition as a function of the anyon mass as it becomes small \cite{mnp}, although the phases do not change the manifestation of the symmetry.    The Chern-Simons term has no effect on the Wilson loop, but it does affect the {\rm '}t Hooft loop.   For a given configuration it adds the linking number of the {\rm '}t Hooft loop with all of the dynamical vortex loops to the action.  We find that both the Wilson loop and the {\rm '}t Hooft loop exhibit a perimeter law even though there are no massless particles in the theory, which is unexpected.  
\end{abstract}

\maketitle
\section{Introduction}
 In this paper, we continue our study of a 3-dimensional lattice loop gas \cite{mnp}, adding a topological interaction which counts the total linking number of the loops with each other.  This Euclidean theory corresponds to a 2+1-dimensional model where the particles and anti-particles are non-interacting hard-sphere anyons.   Thus the only interactions they suffer are a statistical, Aharonov-Bohm \cite{ba} type interaction and an infinite short range repulsion between particles (and anti-particles) which does not allow them to come close to each other.  The Aharonov-Bohm interaction induces a change of phase of the wave function when one particle encircles another, giving a phase dependent on the coefficient of the topological term.  Indeed, it is easy to picture a process whereby two particle anti-particle pairs are created, live for a while and then annihilate.  If the two particles make a 360$^\circ$ turn about one another, the process is described by two loops which link once with each other.   The topological term contributes a phase $e^{i\kappa}$ to the corresponding amplitude, where $\kappa$ is the coefficient of the topological term in the action.  
 
The theory admits an interpretation as an asymptotic, strong coupling, effective Euclidean lattice description of the 2+1 dimensional Abelian Higgs model \cite{mnp} in the symmetry broken sector, the topological interaction corresponding to the addition of a Chern-Simons term (CS), with coefficient $\kappa/2$.  We will lean heavily on our previous work \cite{mnp}; thus we refer the reader to that article for our conventions.  In the symmetry-broken phase, in the asymptotically strong-coupling limit, the theory describes the dynamics of non-interacting vortices and anti-vortices which now behave as non-interacting anyons due to the Chern-Simons term.  The system continues to exhibit the phase transition that was studied in \cite{mnp}.

The Euclidean theory has an action that is no longer real, since the Chern-Simons term adds an imaginary term to the action.  This is in principle an impediment to the simulation of the theory using the Monte Carlo method since the exponential of the Euclidean action cannot be interpreted as a probability density.  However, we can continue to use the real part of the action to give us the probability distribution.  It has the same symmetries as the full action, although it is additionally symmetric under parity.   Then the Chern-Simons term is simply a bounded unimodular phase which can be integrated against the measure that is defined by the real part of the action.
 
 The Chern-Simons term for the configurations that we are left with (non-intersecting closed loops on the lattice) is simply equal to twice the total linking number of all the loops \cite{wz}.  We can see this quite easily by observing that since the Chern-Simons integrand is proportional to the magnetic field, the integral reduces to a set of line integrals along the vortex flux lines:
\bea
S_{CS}&=&\frac{\kappa}{4\pi^2}\int d^3 x\,\, \epsilon^{\mu\nu\lambda}A_\mu\partial_\nu A_\lambda=\frac{\kappa}{8\pi^2}\int d^3 x\,\, \epsilon^{\mu\nu\lambda}A_\mu F_{\nu \lambda}\nonumber\\
&=&\frac{\kappa}{8\pi^2}\left(\sum_{C_i}\oint\,dx^\mu A_\mu\right)\left(\int d^2x_\perp B\right)\nonumber\\
&=&\frac{\kappa}{8\pi^2}\left(\sum_{C_i}2\pi N_{L}(C_i)\right)( 2\pi)=\frac{\kappa}{2}\left(\sum_{C_i} N_{L}(C_i)\right)
\eea
In the second line the integral can be separated into a sum of integrals along each vortex line, denoted $C_i$, and a two-dimensional integral in the transverse direction. The transverse integral is simply equal to the total flux in the vortex line, which is  $2\pi$, while the integral of the gauge field along a given vortex line just gives, via Ampère's law, the total flux of all the other vortex lines that link with the first vortex line.  Thus each integral in the sum is given by $2\pi N_L(C_i)$, where $N_L(C_i)$ is exactly the linking number of vortex line $C_i$ with all of the other vortex lines.   Summing over all the curves clearly gives twice the total linking number of all the vortex loops, $\sum_{C_i}N_L(C_i)=2N_T$, where $N_T$ is the total linking number of the configuration of vortex loops; {\it i.e.},
 \beq
 \frac{\kappa}{2}\left(\sum_{C_i} N_{L}(C_i)\right)=\kappa N_T
 \eeq
 It is this total linking number that we must compute.
 
\section{Computing the linking number}
The closed non-intersecting loops were generated on a body centered cubic (bcc) lattice of size $100^3$ by placing the cube roots of unity randomly on the vertices, with all points on the surface of the lattice assigned the same value.   The lattice can be thought of as filling space with (non-regular) tetrahedra.  Each cube contains six pyramids; adding the diagonal of the cubic sides in a systematic way throughout the lattice divides each pyramid into two non-regular tetrahedra (see \cite{mnp} for details).   If the change in phase of the cube roots of unity around one of the triangular faces of a tetrahedron is equal to $\pm2\pi$, we say a length of vortex flux tube has entered or exited the tetrahedron through that face.  Suppose the flux entered the tetrahedron.  It is a quick exercise to conclude that whatever cube root of unity is placed on the fourth vertex of the tetrahedron, the flux must exit the tetrahedron through one of the other faces.  However then it enters another tetrahedron since the tetrahedra fill space, and the analysis must be repeated.  The loop must close, since the boundary condition used means no surface triangle has a flux passing through it.  The loops so defined exist in the dual lattice to the inital tetrahedral lattice.  

To compute the linking number of any given loop with all of the other loops, we need to simply compute the flux that passes through the given loop, since each other loop that links with it carries one unit of flux.  To compute this flux is actually quite difficult since the given loop and in fact all of the loops exist in the dual lattice where none of the original variables are defined.  On the other hand, the flux that passes through a loop that is defined along the links and vertices of the {\it original lattice} is trivially calculated: we simply calculate the change of phase of the cube roots of unity as we pass through the vertices of such a loop.  Thus if we can systematically deform the given vortex loop on the dual lattice to a loop on the original lattice, the calculation is straightforward.  This deformation is a virtual deformation through space: we do not alter the configuration of the cube roots of unity.  Hence, the calculation of the linking number via the flux that passes through the deformed loop also adds  the linking number of the deformed loop with the original loop, what is reasonably interpreted as a definition for the self-linking number of the original vortex loop.  We must subtract off this self-linking number from the calculation of the linking number of the deformed loop in order to obtain the linking number of the original loop. 

The deformed loop and the original loop form a ribbon, and it is the linking number of the two curves that form the edges of the ribbon that we must calculate.  This is normally called the self-linking number of the original loop.  What we need to do is to subtract off this self-linking number.   The self-linking number of a loop is not a completely well defined quantity \cite{w}: it depends on the deformation that gives rise to the ribbon.  However in this case the deformation is defined by requiring that the deformed loop live on the original lattice.  The self-linking number satisfies the relation \cite{sl}
\beq
\rm Self\,\, linking\,\,  number\,\,  =\,\,  Twist\,\,  +\,\,  Writhe.
\eeq
The Twist of the ribbon is intuitively defined as the number of times the orthonormal frame, defined by tangent vector, the perpendicular displacement vector in the instantaneous plane of the ribbon and the mutually orthogonal vector product of the preceding two vectors, twists around the original curve as you go around the loop.  Surprisingly, the Twist is not an integer.  The Writhe is a more complicated object which also is not an integer, see \cite{writhe} for details.  It is intuitively the number of coils without twist that the ribbon suffers.  Of course the sum of the two is an integer, the self-linking number of the ribbon.  A familiar example might correspond to a coiled, untwisted extension cord.  When we straighten it out, invariably it becomes all twisted up; this is simply because writhe is converted into twist, as the self-linking number is essentially invariant.  

However, the self-linking number is in fact the ordinary linking number of the deformed loop with the original loop.  The Gauss formula for the linking number is given as two line integrals over the locations of the two loops which link:
\beq
N_{SL}=\frac{1}{4\pi}\oint_{{\cal C}_1}\oint_{{\cal C}_2}\frac{d\vec x\cdot(d\vec y\times(\vec x-\vec y))}{|\vec x -\vec y|^3},
\eeq
where $d\vec x$ and $d\vec y$ are the line elements along the two loops respectively.  These integrals are prohibitively time consuming to calculate directly since the loops are polygonal with of the order of $10^6$ segments.  Hence one is left with about $10^{12}$ integrals to do, which, although not impossible, is not feasible 
with our computing resources.  However there is a simpler way to compute this linking number, using knot theory \cite{sl}.  

\begin{figure}[ht]
 \centerline{\includegraphics[width=\linewidth]{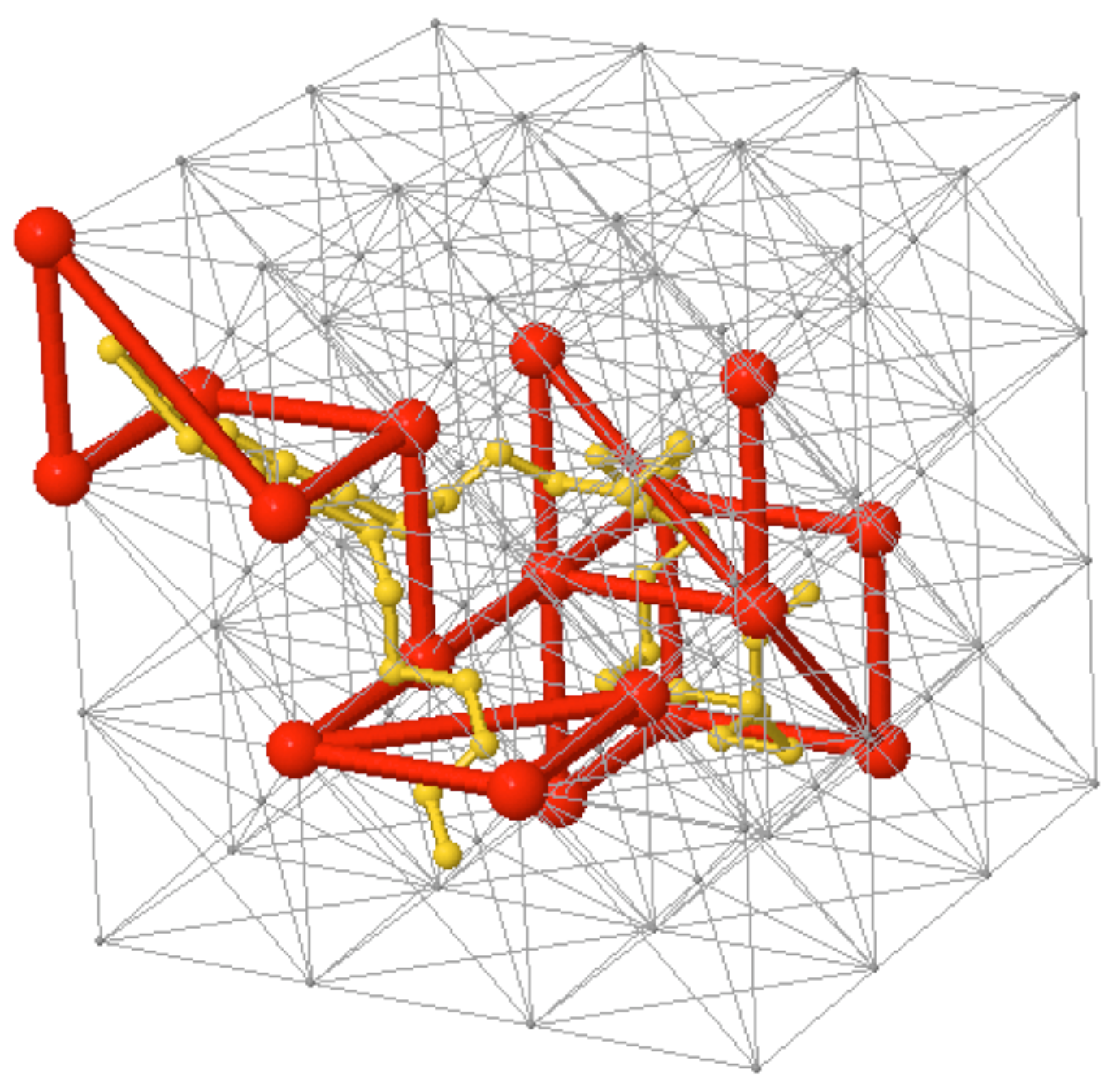}} \caption{\label{pqc} (color online) An example of a vortex loop (thin, yellow) and its deformation to the lattice (thick, red) viewed from a random direction.}
\end{figure}

The idea is the following.  We can project the knot onto a two dimensional plane, keeping track only of the the sense of the crossings of the segments of one loop with the other in the projection.  Then a simple sum of the association of $\pm 1$ to the crossings, depending on which segment of which loop is on top of the other and its direction, gives the linking number.  The important point is to choose the direction of projection that will yield the most simplifying two dimensional projection.  Indeed, an arbitrary projection will yield just as many independent crossings as there are links in the two loops, not significantly reducing the number of calculations (see Fig. \ref{pqc}).  

\begin{figure}[ht]
 \centerline{\includegraphics[width=\linewidth]{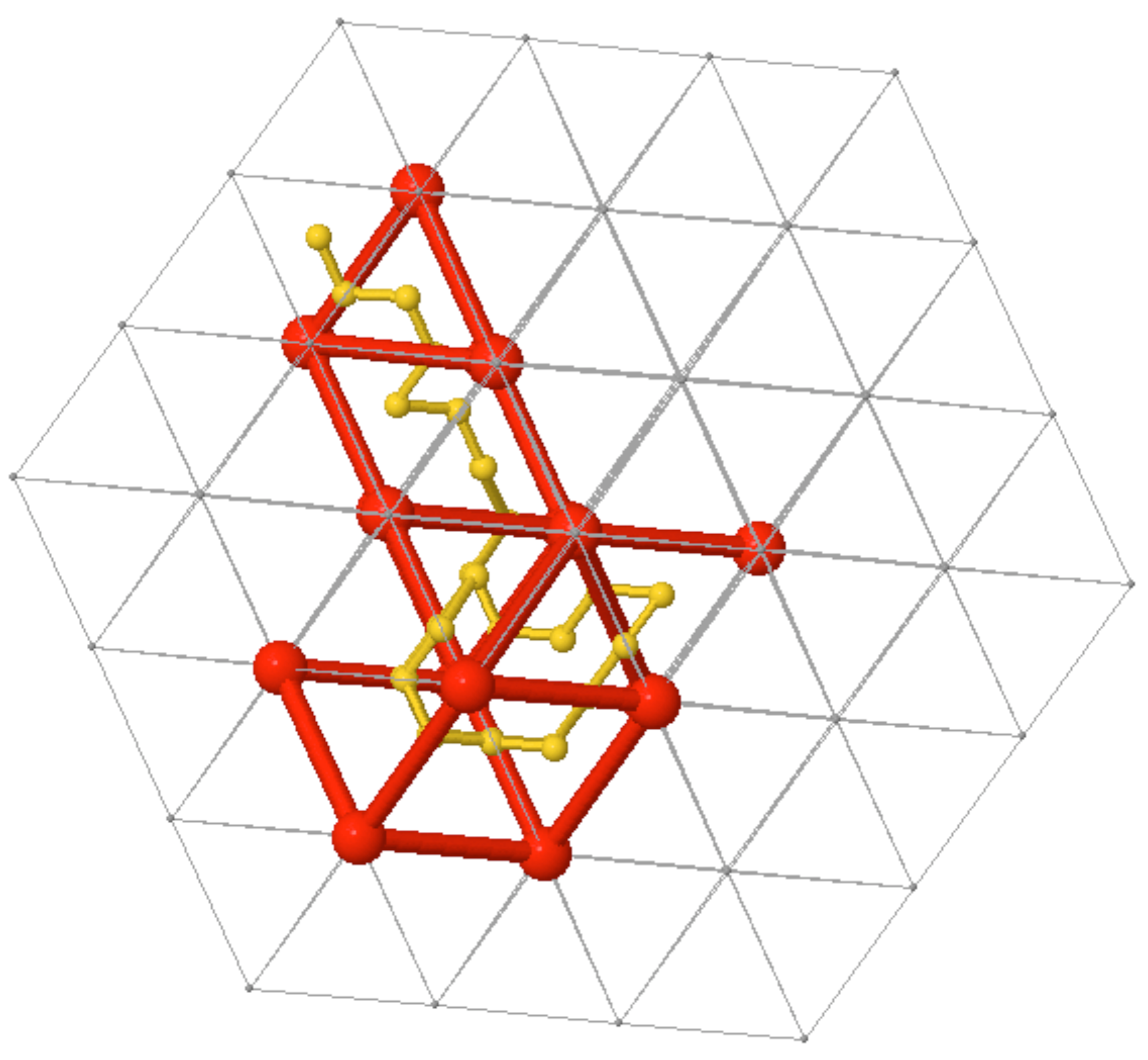}} \caption{\label{p111} (color online) The same vortex loop and its deformation as in Fig. \ref{pqc} viewed along the $(1,1,1)$ direction.}
\end{figure}

However, projecting along the diagonal of the lattice (the $(1,1,1)$ direction in coordinate space) actually yields exactly a regular triangular lattice on the projected two plane.  The original loop, which passes through the dual lattice, projects to the dual lattice of the two dimensional triangular lattice, while the deformed loop, of course, projects directly to the links of the triangular lattice (see Fig \ref{p111}).  

Thus the crossings are unambiguous and occur at a small, finite number of intersection points.  It is easy to keep track of the segments of each loop, and their relative heights.   This simply amounts to a re-indexing of the data which is already stored in the computer, in a new system of coordinates given by the triangular lattice in the two dimensional projected plane and the height along the $(1,1,1)$ direction.  The original calculation of the linking number which would take several days of computer time reduces to a few seconds.  

\section{Calculating in the Chern-Simons Theory}
The Monte Carlo process for generating a set of configurations in the presence of the Chern-Simons term is not straightforward.   The point is that the Chern-Simons term adds an imaginary contribution to the Euclidean action; hence it cannot be used to define the probability distribution required to obtain the equilibrium configurations via the Monte Carlo method.   However, we obtain the set of equilibrium configurations by using the Boltzmann weight given by the total length of the vortex loops.  Thus the Boltzmann weight is unaffected by the Chern-Simons term.

This is a classic problem for numerical simulations; it arises in theories with fermions where it is called the sign problem, but also in the context of topological terms which are odd under time reversal; a fuller explanantion of this is provided, for example, in \cite{ampr}.  The solution in the present case is based on the following logic.  The (Euclidean) Feynman path integral instructs us to integrate 
\beq
\langle{\cal O}\rangle=\int {\cal D}\varphi\,\, e^{-S_E/\hbar}{\cal O}
\eeq 
over the space of field configurations, with appropriate boundary conditions,  to calculate  a quantum amplitude $\langle{\cal O}\rangle$.  However, the actual measure on this space is not specified.  There does not exist a naive, canonical, translationally invariant (translations in the space of functions) measure on the space of functions \cite{gj} that we can use.  However perfectly well defined Gaussian measures do exist, and are defined by quadratic (free) field theories.  It is the problem of constructive quantum field theory to prove that the set of amplitudes created by the Feynman path integral using the Gaussian measure on function space, integrating the exponential of the interactions in the Lagrangian against this measure, provides a full set of finite Greens functions that satisfy the Wightman axioms \cite{wi}, which then allow for a reconstruction of the corresponding, interacting quantum field theory.  In the present case, the interaction could indeed comprise of the Chern-Simons term, which we note, remains imaginary in Euclidean space.  The Gaussian measure suffices to define the quantum field theory, modulo some amount of (infinite) renormalisations of coupling constants, however, the net effect of the construction of the quantum field theory is in fact to define a measure corresponding to the exponential of the full action, not just the Gaussian part. There are several examples of this kind of construction of non-trivial quantum field theory \cite{s} and the corresponding measures.   

The important property that the measure, used to construct the quantum field theory, must satisfy is that it respect all the symmetries of the theory that one is trying to construct.  In our case, we do not use only the Gaussian part of the action, since we are doing a non-perturbative, lattice based numerical analysis.  We use the full real part of the action with the approximation that yields, for the action, the total length of the vortex loops for the field configurations that we consider.  The full real part of the action actually has more symmetry than the theory with the Chern-Simons term added since parity and time reversal are not preserved.  However this does not cause any problem since all amplitudes are calculated with the Chern-Simons term inserted.   

Hence we use the full real part of the action  to define the measure on the space of field configurations, and in the context of the Monte-Carlo method, the set of equilibrium configurations, say a total number $N$.  After this, the Chern-Simons term simply gives a uni-modular phase that can be integrated against this measure.  When calculating actual matrix elements of an operator, we must calculate the average of the operator with the phase coming from the Chern-Simons term inserted and then divide by the partition function, again defined with the same phase inserted.  Explicitly we get
\beq
\langle {\cal O}\rangle \approx \langle {\cal O}\rangle_N=\frac{\sum_{i=1}^N {\cal O}({\cal C}_i) e^{iS_{CS}({\cal C}_i) }}{\sum_{i=1}^N  e^{iS_{CS}({\cal C}_i) }}\label{e3}
\eeq
where ${\cal C}_i$ stands for the $i$th configuration.  Normally, both numerator and denominator contain a factor of $1/N$; however, here it cancels between them.

\subsection{Chern-Simons term}
\begin{figure}[ht]
 \centerline{\includegraphics[width=\linewidth]{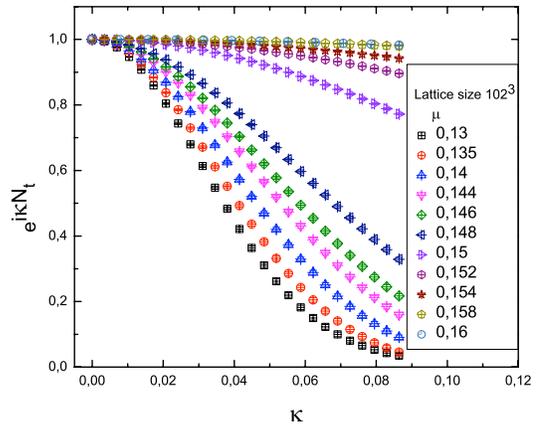}} \caption{\label{f2} (color online) The average value of the Chern-Simons term  as a function of $\kappa$.}
\end{figure}

The Chern-Simons term in the functional integral gives exactly the linking number of all the dynamical loops.  We use knot theoretic techniques to compute the average of the Chern-Simons term in the set of equilibrium configurations, for different values of the mass $\mu$.  We see that the expectation value drops to zero remarkably quickly as the system passes through the transition at about $\mu =0.152$ for decreasing $\mu$.  
The graph for the expectation value of the Chern-Simons term is given in Fig. \ref{f2} for various values of $\mu$.
This average of the Chern-Simons term serves as the partition function when computing the expectation value of any operator.  Clearly proceeding to values of $\kappa\gtrsim 0.08$ is not possible for $\mu \lesssim 0.15$.

\subsection{Wilson loop}
The Wilson loop \cite{wilson} is defined as the expectation value of the operator
\beq
W= e^{-i(q/e)\oint A_\mu dx^\mu}
\eeq
where $e$ is the fundamental charge in the model and the integral in the exponent goes along a closed, fixed rectangular path of width $L$ and length $T$.  For a given configuration ${\cal C}_i$, the exponent in the Wilson loop is equal to the linking number of the curve defining the Wilson loop and the dynamical vortex loops in the configuration, say $N_{WL} ({\cal C}_i)$,
\beq
W({\cal C}_i)=e^{-i(2\pi q/e) N_{WL}({\cal C}_i)}.
\eeq

The calculation of the expectation value of the Wilson loop can be done by first calculating the average value of the Wilson loop for fixed total linking number, and then performing the sum over these average values, weighted by the number of configurations with the fixed total linking number.  Using the notation ${\cal N}(N_T)$ for the number of configurations with fixed total linking number $N_T$, and ${\cal C}_{i, N_T}$ as an index for these configurations, we have:
\begin{eqnarray}
\langle W \rangle &=& \frac{
\sum_i W({\cal C}_i) e^{i\kappa N_T({\cal C}_i )}}
{\sum_i e^{i\kappa N_T({\cal C}_i )}}\cr
&=&\frac{
\sum_{N_T} {\cal N}(N_T) \left(\frac{1}{{\cal N}(N_T)}\sum_{{\cal C}_{i, N_T}} W({\cal C}_{i, N_T})\right) e^{i\kappa N_T}}
{\sum_{N_T}  {\cal N}(N_T) e^{i\kappa N_T}}.
\end{eqnarray}

\begin{figure}[ht]
 \centerline{\includegraphics[width=\linewidth]{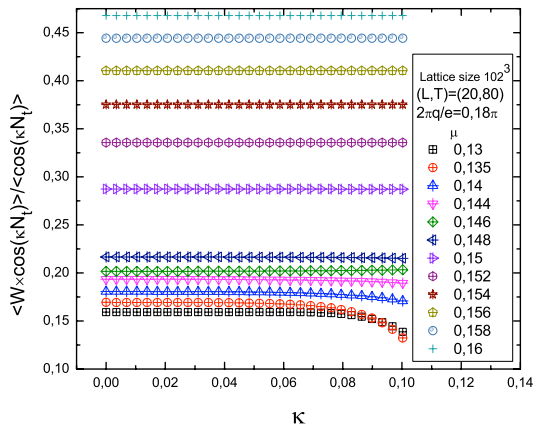}} \caption{\label{f1} (color online) The Wilson loop for various values of the coefficient of the Chern-Simons term:  $\kappa$.}
\end{figure}

The term in parentheses in the numerator is the average value of the Wilson loop with fixed total linking number. If this is independent of the value of the total linking number, then it comes out of the sum, and in fact the sums in the numerator and denominator cancel, yielding
\beq
\langle W \rangle=\frac{1}{{\cal N}(N_T)}\sum_{{\cal C}_{i, N_T}} W({\cal C}_{i, N_t})
\eeq
which is in fact independent of $\kappa$.  This is exactly what we find with our numerical simulation.   In Fig. \ref{f1} we plot the value of the Wilson loop for different values of $\mu$ as a function of $\kappa$, at a fixed value $2\pi q/e=0.18\pi$.  Evidently, the average of the Wilson loop does not depend on $\kappa$ for any value of $\mu$ for $\kappa \lesssim 0.08$.   

There is an apparent dependence in the graphs for small $\mu$, as $\kappa$ exceeds the value $\sim 0.08$.  However, at this point, as we can see from Fig. \ref{f2} the average value of the Chern-Simons term, by which we divide, becomes very small, and we no longer trust the numerical results.  Indeed, for small $\mu$, the number of configurations at the extremities of the distribution of the total linking number ({\it i.e.} at large total linking number, which only occurs for small $\mu$) becomes only a handful.  Then the average of the Wilson loop operator for these configurations deviates wildly with respect to the average when there are many configurations ({\it i.e.} about 100) deviations which are magnified when the denominator also becomes small. We have verified that if we increase the number of configurations that we have available at fixed total linking number, then the average of the Wilson loop for this set of configurations converges to the $\kappa$ independent value as the number of configurations becomes large $\sim 100$.  

\subsection{{\rm '}t Hooft loop}
\begin{figure}[ht]
 \centerline{\includegraphics[width=\linewidth]{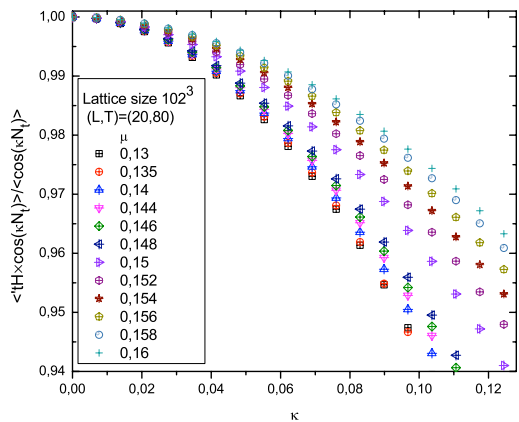}} \caption{\label{f3} (color online) The average value of the {\rm '}t Hooft loop in the presence of a Chern-Simons term  as a function of $\kappa$.}
\end{figure}

The {\rm '}t Hooft loop \cite{thooft} corresponds to the insertion of a singular magnetic flux tube along a contour of a fixed rectangular loop of width $L$ and length $T$.  It is the dual object to the Wilson loop \cite{gw}.  It is important to note that this is not a vortex loop, but just a gauge field loop.  Thus the functional integral over the gauge fields is subject to the constraint that such a magnetic flux loop exists at the given fixed position. The Monte Carlo method of generating the equilibrium configurations is unchanged, using as before only the real part of the full action, with the (infinite) action of the {\rm '}t Hooft loop subtracted off and with our strong coupling approximation.  Then the equilibrium configurations are comprised of configurations of closed vortex loops appended by the {\rm '}t Hooft loop.      In the presence of the Chern-Simons term, the {\rm '}t Hooft loop simply adds $\kappa N_{{\rm '}t HL}$ to the action, where $N_{{\rm '}t HL}$ is the linking number of the {\rm '}t Hooft loop with all the dynamical vortex loops.  Hence the {\rm '}t Hooft loop is given by the average
\beq
\langle {\rm '}t H\rangle = \frac{\sum_{{\cal C}_i}e^{i\kappa N_{{\rm '}t H}}e^{-S_E+i\kappa N_T}}{\sum_{{\cal C}_i}e^{-S_E+i\kappa N_T}}
\eeq
In Fig. \ref{f3} we plot the average value of the {\rm '}t Hooft loop as a function of $\kappa$, for various values of the mass $\mu$.  The points in the graphs beyond $\kappa = 0.08$ should not be trusted for small values of $\mu$, since the errors are not under control.  The average value of the Chern-Simons term in the denominator, as we apply Eqn. (\ref{e3}), becomes vanishingly small.  We note that in contrast with the Wilson loop, the {\rm '}t Hooft loop has a clear dependence on the coefficient of the Chern-Simons term.  The 't Hooft loop is constant (equal to 1) in the absence of the Chern-Simons term, but is a function of the coefficient of the Chern-Simons term, in its presence.
 
 \maketitle
\section{ DISCUSSION AND CONCLUSIONS }
The main result that we find is that the Wilson loop \cite{wilson} is independent of the Chern-Simons term while  the {\rm '}t Hooft loop \cite{thooft} is not.  Both are expressed as the exponential of a contribution to the free energy.  In fact for the {\rm '}t Hooft loop the contribution to the free energy vanishes in the absence of the Chern-Simons term but is proportional to its coefficient in its presence.  The free energy for the Wilson loop is simply independent of the Chern-Simons term.  

These results are already surprising; however, there is an even more astonishing result.  The fact is that both the Wilson loop and the {\rm '}t Hooft loop have perimeter law behaviour in both phases of the theory.  This is even more remarkable since there is an understanding, at least in $SU(N)$ (non-abelian) and in $Z_N$ (abelian) gauge theories   that a perimeter law for both of these order parameters requires the existence of massless particles \cite{gw}.  It is not unreasonable to believe that non-compact $U(1)$ gauge theory that we study, will behave in the same way, especially given that the large $N$ limit of $Z_N$ gauge theory does give compact $U(1)$ gauge theory.  However, we have no massless particles in the theory.  We believe that we manage to circumvent the previous conclusion since we have a statistical long range interaction between the anyons.  

To see that the order parameters do indeed exhibit a perimeter law, we refer to our previous article \cite{mnp}.  In this article it was established, through detailed numerical analysis, that the average of the Wilson loop exhibits a perimeter law as a function of its parameters $L$ and $T$. It was also found that it exhibits this behaviour on either side of the phase transition, which occurs at around $\mu\approx 0.152$.  However the Wilson loop average is calculated from the linking number of the fixed loop of dimension $L$ and $T$ with the dynamical vortex loops.  But the expectation value  of the {\rm '}t Hooft loop is based on exactly the same calculation.  For the Wilson loop we compute the average value 
\beq
\langle W \rangle = \frac{\sum_{{\cal C}_i}e^{-i(2\pi q/e)N_{WL}}e^{-S_E+i\kappa N_T}}{\sum_{{\cal C}_i}e^{-S_E+i\kappa N_T}}
\eeq
while the {\rm '}t Hooft loop we compute
\beq
\langle {\rm '}t H \rangle = \frac{\sum_{{\cal C}_i}e^{i\kappa N_{{\rm '}t H}}e^{-S_E+i\kappa N_T}}{\sum_{{\cal C}_i}e^{-S_E+i\kappa N_T}}.
\eeq
However for a given rectangular loop, $N_{WL}=N_{'tH}$; hence, the perimeter behaviour that was established for the Wilson loop in \cite{mnp} goes over to the 't Hooft loop.  

\maketitle
\section{ACKNOWLEDGEMENTS}
We thank NSERC of Canada for financial support, the African Institute for Mathematical Sciences (AIMS) for hospitality,  where much of this paper was written up, and the Réseau québécois de calcul de haute performance (RQCHP) for providing us with the computational resources required for this work.


\end{document}